\documentclass[
    twocolumn,
	prd,
	amssymb,
	preprintnumbers,superscriptaddress,
	nofootinbib]{revtex4-1}

\pdfoutput=1

\usepackage{graphicx}
\usepackage{enumitem}
\usepackage{latexsym}
\usepackage{amsfonts}
\usepackage{amssymb}
\usepackage{xcolor}
\usepackage[export]{adjustbox}
\usepackage{amsmath}
\usepackage{slashed}
\usepackage{dcolumn}
\usepackage{verbatim}
\usepackage{float}
\usepackage{multirow}
\usepackage{xspace}
\usepackage[normalem]{ulem}
\usepackage[
pdfauthor={Djuna Croon}]{hyperref}

\newcommand{\beq}{\begin{equation}}
\newcommand{\eeq}{\end{equation}}
\newcommand{\bea}{\begin{eqnarray}}
\newcommand{\eea}{\end{eqnarray}}

\allowdisplaybreaks

\begin{document}
\title{Light Dark Matter through Resonance Scanning}
\author{Djuna Croon}
\email{dcroon@triumf.ca}
\affiliation{TRIUMF Theory Group, 4004 Wesbrook Mall, Vancouver, B.C. V6T2A3, Canada}

\author{Gilly Elor}
\email{gelor@uw.edu}
\affiliation{Department of Physics, University of Washington, Seattle, WA 98195, USA}

\author{Rachel Houtz}
\email{rachel.houtz@durham.ac.uk}
\affiliation{Institute for Particle Physics Phenomenology, Department of Physics, Durham University, Durham DH1 3LE, U.K.}

\author{Hitoshi Murayama}
\email{hitoshi@berkeley.edu, hitoshi.murayama@ipmu.jp}
\affiliation{ Berkeley Center for Theoretical Physics, University of California, Berkeley, CA 94720, USA}
\affiliation{ Theory Group, Lawrence Berkeley National Laboratory, Berkeley, CA 94720, USA}
\affiliation{ Kavli IPMU (WPI), UTIAS, The University of Tokyo, Kashiwa, Chiba 277-8583, Japan}
\author{Graham White}
\email{graham.white@ipmu.jp}
\affiliation{ Kavli IPMU (WPI), UTIAS, The University of Tokyo, Kashiwa, Chiba 277-8583, Japan}

\preprint{}

\begin{abstract}
We propose a new out-of-equilibrium production mechanism of light dark matter: resonance scanning. If the dark matter mass evolved in the early Universe, resonant production may have occurred for a wide range of light dark matter masses today.
We show that the dark matter relic abundance may be produced through the Higgs portal, in a manner consistent with current experimental constraints.
\end{abstract}

\maketitle


\section{Introduction}
Light dark matter has become a focus of the next generation of direct-detection experiments \cite{Hochberg:2017wce,Griffin:2018bjn,Hochberg:2019cyy,Kurinsky:2019pgb,Coskuner:2019odd,Trickle:2019ovy,Griffin:2020lgd,Knapen:2017ekk}. However, unlike for the popular weakly interacting massive particles (WIMP), thermal production of sub-GeV dark matter requires a cross section in tension with existing cosmological constraints. 
As such, production mechanisms for dark matter in this mass range are limited \cite{Hall:2009bx,Hochberg:2014kqa,Dvorkin:2020xga}.\par

In this work we propose a new mechanism for the production of light dark matter, through a Higgs portal. The mechanism is based on the hypothesis that the dark matter mass was not constant in the early Universe, but instead depends on the vacuum expectation value (VEV) of an ultralight scalar, the \emph{morphon}. As the Universe cools, the temperature-dependent morphon VEV implies a decreasing dark matter mass. 
The scanning dark matter mass will generically pass through $m_\chi \approx m_h/2$, implying resonant enhancement of Higgs portal production. In this way, light dark matter can be produced through a smaller portal coupling $\lambda_{h{\rm DM}}$ than in the standard case. 
The dark matter only obtains its final sub-GeV mass \emph{after} losing contact with the thermal bath.

The mechanism opens up the parameter space of existing Higgs portal dark matter models ~\cite{Davoudiasl:2004be,Patt:2006fw,Athron:2018hpc,Beniwal:2019xop,Arcadi:2019lka}. 
The parameter space of the standard Higgs portal WIMP scenario has been encroached on by direct detection and collider experiments~\cite{Goodman:2010ku,Carpenter:2013xra,Petrov:2013nia,Berlin:2014cfa,Abercrombie:2015wmb,Amole:2015pla,Aprile:2016swn,Akerib:2016lao,Fu:2016ega,Fu:2016ega,Amole:2016pye,Escudero:2016gzx,Ghorbani:2016edw,Akerib:2016vxi,Sirunyan:2017hnk,Aaboud:2017phn,Aaboud:2017bja,Aaboud:2017yqz,Cui:2017nnn,Aprile:2019dbj,Aprile:2017iyp}. In fact, the only remaining allowed parameter space corresponds to the special case that the dark matter mass is near half of the Higgs mass, such that its interactions with the SM through the Higgs portal are resonantly enhanced. The mechanism presented here implies that this enhancement is not coincidental, but a generic stage in the thermal evolution of Higgs portal dark matter.

A specific example of the resonance scanning mechanism is given by a non-minimally coupled morphon field. The evolution of the scalar curvature, in particular through the anomalous breaking of scale invariance near the QCD scale, gives the morphon a varying potential and the dark matter a falling mass. We will demonstrate the success of this scenario in detail. 

Dark matter with a dynamical mass has previously been considered as a way to open/close new production channels \cite{Baker:2016xzo,Baker:2017zwx,Baker:2018vos,Bian:2018bxr}, dynamincally trigger freezeout~\cite{Heurtier:2019beu}, raise the dark matter mass observed today to explain an anomaly \cite{Cohen:2008nb}, make use of an exotic confinement transition \cite{Croon:2019ugf} or make dark matter heavier than is normally allowed by perturbativity \cite{Davoudiasl:2019xeb}. 
This work complements these existing proposals by decoupling the dark matter mass today from the dark matter mass through which the yield was set, allowing a range of masses to benefit from resonantly enhanced production.

\section{Resonant production}
Let us first review some key principles of resonant production in Higgs portal models. Throughout this work, we assume the dark matter is a Dirac fermion, below the weak scale defined by
\begin{align}
    \mathcal{L}_\chi &=   \bar{\chi}\gamma^\mu\partial_\mu\chi-m_{\chi}\bar{\chi}\chi - \frac12 \lambda_{h \chi} h \bar\chi \chi \ , 
\end{align} 
where $m_{\chi}\sim$ MeV is the $T=0$ mass of the dark matter fermion. The last operator in $\mathcal{L}_\chi$ -- the portal coupling to the SM Higgs field -- breaks SU(2)$_L$ and therefore $\lambda_{h \chi} \propto v/\Lambda$ (where $\Lambda$ is a new physics scale) may be small.

In standard thermal freeze-out, the portal coupling controls how much dark matter annihilates before it decouples, and therefore sets the relic abundance. Alternatively, in freeze-in scenarios the portal coupling controls the production of dark matter though the same diagram. The dark matter annihilation into SM final states proceeds through a Higgs boson in the s-channel.   When the dark matter mass is in the region $m_\chi \simeq m_h/2$, the annihilation cross section is resonantly enhanced.

Direct detection experiments~\cite{Aprile:2017iyp,Aprile:2019dbj} and current limits on invisible Higgs decays~\cite{Aad:2015pla} only allow for small portal couplings $\lambda_{h\chi}$. On the other hand, the annihilation cross section must be large enough that freeze-out reproduces the correct relic abundance. In order to simultaneously achieve relic abundance and avoid direct detection bounds, the dark matter must annihilate through the Higgs mediator on resonance. The enhanced annihilation cross section allows for comparatively small Higgs portal couplings. This effect can be seen in the line of the relic abundance curve for Higgs portal models, reproduced here as the dark blue curve in Fig.~\ref{fig:relic}. The resonance peak appears instead as a valley on this plot, and the relic abundance curve dips into small enough couplings to avoid Xenon1T bounds just as $m_\chi$ approaches $m_h/2$, highlighted by the dotted vertical line.

\section{Dark matter mass morphing}

Here we will consider coupling the dark matter to a pseudoscalar \emph{morphon} field $\phi$,
\begin{align}
    \mathcal{L} &= \mathcal{L}_{\rm SM} + \mathcal{L}_{\chi} - i y_{\phi \chi} \phi \bar\chi \gamma_5 \chi- V(\phi)
\end{align}
We will study the scenario in which the morphon potential $V(\phi)$ implies a temperature dependent evolution of the morphon vacuum expectation value $v_\phi(T)$, such that the morphon field contributes a temperature dependent piece to the effective dark matter mass,
\begin{align}
    m_{\chi }^2 (T) &= \left[ m_{\chi,0} + \lambda_{h \chi} v_h \right]^2+ y_{\phi\chi}^2 v_\phi^2(T)
    \label{eq:Lag}
\end{align}
where the terms within brackets indicate the zero temperature mass which we will simply refer to as $m_\chi$. 

Resonance scanning relies on the evolution of the morphon field $\phi$ from a nonzero field value ($v_\phi(T)>0$) to a zero field value ($v_\phi(T)=0$) at low temperatures. 
If the shift in dark matter mass is significant, the morphon potential has to be relatively flat, if it is not to contribute significantly to the vacuum energy at early times. It would therefore be natural to consider scenarios in which the morphon field is a pseudo-Goldstone boson, in which case \eqref{eq:Lag} can be considered an expansion around the minimum of the potential.

As a proof of concept, we consider a scenario in which the morphon field is coupled to the Ricci-scalar, 
    \begin{equation}
        V(\phi)= \frac12 m_\phi^2 \phi^2 - \frac{1}{2} \xi R \phi^2 + \frac14 \lambda_\phi \phi^4 \,, 
    \end{equation}
     where we choose $\xi  = 1/6$, motivated by conformal invariance \cite{Sonego:1993fw, Dowker:1970vu}. Such a coupling has earlier been studied in \cite{Opferkuch:2019zbd,Davoudiasl:2004gf}.
The temperature evolution of the Ricci scalar generates a temperature-dependent mass term in the potential. If the equation of state were constant, the Ricci scalar would be $(3w-1)H^2$, which may lead one to believe that $R$ vanishes during radiation domination. Anomalous effects, however, reintroduce $R\neq 0$ in this regime. The Ricci scalar is proportional to the trace of the stress-energy tensor $T^\mu_\mu$, such that it is non-vanishing due to the anomalous breaking of scale invariance (the trace anomaly, $T^\mu_\mu\neq0$). The effects that break scale invariance during the relevant period of the morphon's phase transition are the perturbative running of couplings, mass thresholds, and QCD confinement. 

Estimating the size of $R$ from the running of the couplings only, we find $R/H^2 \sim 10^{-2}$--$10^{-1}$ for $T\sim$ GeV, with a quick increase near QCD confinement where the perturbative analysis breaks down~\cite{Davoudiasl:2004gf}. The full analysis, including a careful treatment of confinement and deviations from scale invariance due to mass thresholds, has been done in~\cite{Caldwell:2013mox}. In our calculation, we will use their result for the evolution of the Ricci scalar between $T\sim 100$ MeV and $T\sim 10$ GeV. This is plotted as $\sqrt{R(T)}$ in Fig.~\ref{fig:rates}.
    
    At high temperatures the morphon develops a nonzero VEV:
    \begin{align}
        v_\phi (T) = 
            \left\{
            \begin{array}{cl}
            \sqrt{ \frac{1}{\lambda_\phi} \left( \xi R - m_\phi^2 \right)} \,, \ \ \ 
                     &    \xi  R  > m_\phi^2 \\[11pt]
            0       & \text{otherwise}\,.
            \end{array}
            \right. 
    \end{align}
In this way, the morphon VEV decreases with decreasing $T$, until it vanishes at low temperatures. The dark matter loses the extra contributions to its mass from interactions with the morphon, realizing its final light mass. The red solid line of Fig.~\ref{fig:rates} shows the temperature evolution of the dark matter mass for a sample benchmark point. The dark matter mass triggers the resonance when it is equal to half of the Higgs mass, $m_\chi(T_{\rm res})=m_h/2$.

\section{Consistent Thermal averaging}
In order to study the production of dark matter in this scenario, we will solve a Boltzmann equation of the form,\footnote{Typically the decay rate $\Gamma_{H\rightarrow \chi\chi}$ dominates for $m_\chi \ll m_h$. However, for resonance scanning dark matter this decay channel only becomes kinematically accessible below the resonance temperature. For $T_{\rm res} \sim$ GeV, we find $n_{h,{\rm eq}}(T_{\rm res}) \ll n_{\chi, {\rm eq}}(T_{\rm res})$, an inequality that grows rapidly as $m_\chi$ further falls with the temperature.
At temperatures below $T_{\rm res}$, we find $\Gamma_{H\rightarrow \chi\chi}n_{h,{\rm eq}} \ll \langle \sigma_{\bar\chi\chi} v \rangle n^2_{\chi, {\rm eq}}$ and will therefore focus on the annihilation rate here. } 
    \begin{align}\label{eq:BE}
        \frac{dn}{dt} + 3 H n
            &= - \left(n^2 - n_{eq}^2 \right) \langle \sigma_{\bar\chi\chi} v \rangle \
    \end{align}
The simple form of this equation relies on thermally averaging the dark matter annihilation cross section, a procedure based on the momentum distribution functions for the dark matter particles. If the  mass changes more rapidly than the rates establishing kinetic equilibrium, the distribution functions are no longer well-defined. Therefore, we introduce three conditions to ensure the formalism is appropriate, 
\begin{enumerate} 
    \item Kinetic equilibrium, \begin{equation} \label{eq:kineq}
        \Gamma_{\rm scatt}> H\,. \end{equation}
    \item The reactions that establish kinetic equilibrium are faster than the rate of change of the mass, 
        \begin{equation} \label{eq:masscond}
            \Gamma_{\rm scatt} > \frac{\dot{m}}{m}\,.
        \end{equation}
    \item The adiabatic condition, 
        \begin{equation} \label{eq:adiabatic}
            \frac{d m_\chi}{dt} < m_\chi^2
        \end{equation} 
        ensures that non-perturbative dark matter production is not significant.
    \end{enumerate}
    
\begin{figure}[t!]
    \centering
    \includegraphics[width=0.5\textwidth]{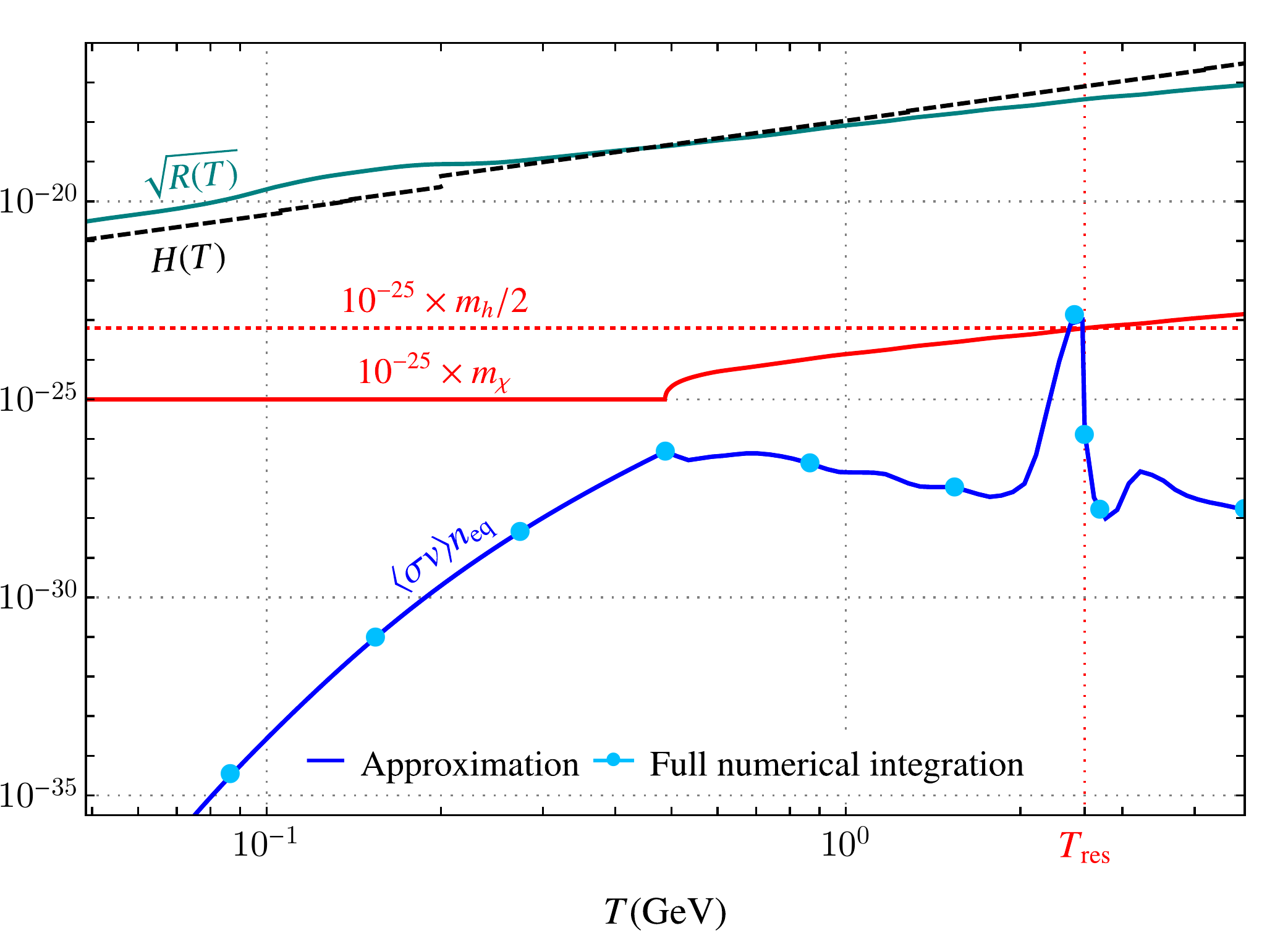}
    \caption{Relevant rates for a benchmark with $m_\chi = 1$ GeV, $m_\phi = 10^{-10}$ eV and $\sqrt{\lambda_\phi} = 2.5 \times 10^{-20} y_{\phi \chi} $. The masses $m_\chi$ and $m_h/2$ are rescaled to fit in the same plot. All quantities are quoted with GeV unit. 
    }
    \label{fig:rates}
\end{figure}

It may naively seem like there is very little parameter space which passes the first two conditions, without bringing the DM into thermal equilibrium. This is indeed the case for the scenario in which DM is bosonic. However, for fermionic DM, the annihilation cross section is p-wave suppressed with respect to the scattering cross section. This motivates our choice to focus on fermionic DM: as we will see, it is possible to find parameter space which produces DM out-of-equilibrium while also passing these checks.

\begin{figure}[t!]
    \centering
    \includegraphics[width=0.5\textwidth]{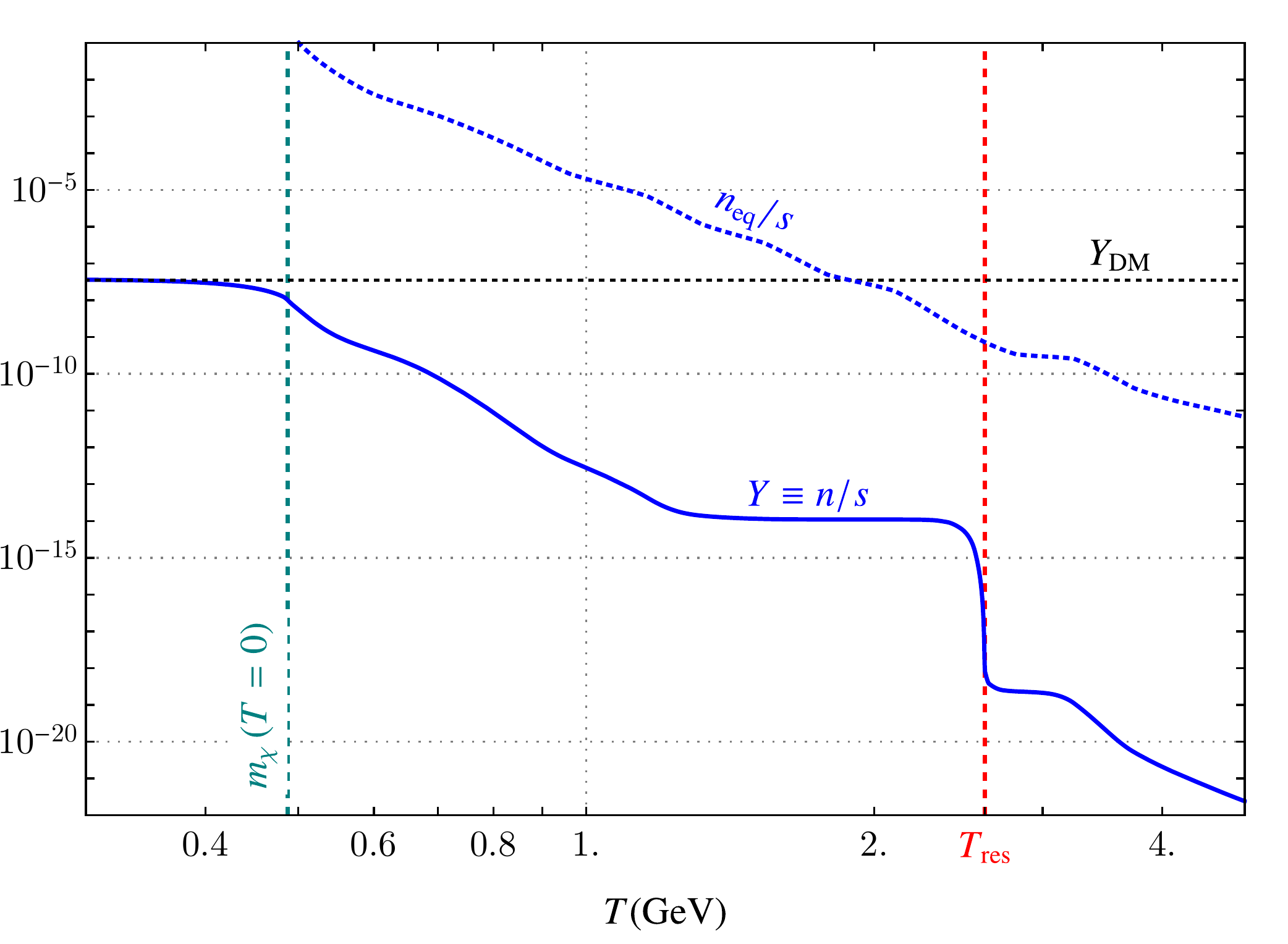}
    \caption{Evolution of the yield for a benchmark with $m_\chi = 10$ MeV, $m_\phi = 10^{-10}$ eV and $\sqrt{\lambda_\phi} = 2.5 \times 10^{-20} y_{\phi \chi} $. Here the yield is defined with respect to the entropy density today $s_0$, and rescaled by $(T/T_0)^3$.
    }
    \label{fig:yield}
\end{figure}

In the region where we trust our thermal averaging prescription, we are free to use the expression
    \begin{align}
    \langle \sigma v \rangle
		&=  \int_{4 m_\chi^2 }^\infty{ds 
		    \frac{ s \sqrt{ s - 4 m_\chi^2 } K_1\left( \frac{ \sqrt s }T \right) \sigma v_{rel}^{cms} }
	    	{16 T m_\chi^4 K_2^2\left( \frac{ m_\chi }T \right)}} \ ,
	\label{eq:thermal-ave}
    \end{align}
where the annihilation cross section through the Higgs portal is given by
    \begin{align}
    \sigma v_\text{rel}^\text{cms}
        &= P_\chi \frac{ 2 \lambda_{h\chi}^2 v_h^2 }{ \sqrt s }
        \frac{ \Gamma_h ( m_h^* = \sqrt s ) }
            { (s - m_h^2 )^2 + m_h^2 \Gamma_h^2 (m_h ) } \ , 
\\
    P_\chi &=  \frac s 2 \left( 1 - \frac{ 4 m_\chi^2 }s \right) \,,
    \end{align}
following the notation of~\cite{Athron:2018hpc}.

Around the resonance peak, thermal averaging constitutes a numerical challenge due to the rapid variation of the cross section. In this regime we utilize a change of integration variables
    \begin{align}
    s = m_h^2 + m_h \Gamma_h \tan\eta \ ,
    \end{align}
which renders the integrand in Eq.~\eqref{eq:thermal-ave} flat. 
Away from the resonance peak,  we switch back to integration in $s$ and the smooth, non-resonant behavior leaves us free to use sampling and interpolation to speed up the thermal averaging integration. We demonstrate the success of this approach in Fig.~\ref{fig:rates} by a direct comparison with a numerical integration at machine precision.

\section{Relic abundance}
We solve the Boltzmann equations in a radiation dominated era, $t = 1/2H$,  which implies that we can rewrite the standard Boltzmann equation \eqref{eq:BE} as
\begin{equation}
 \frac{d \log n}{d \log T} - 3 = \frac{1}{ H} \langle \sigma v \rangle \left( 10^{\log n} -  \frac{n_{\rm eq }^2}{10^{\log n}}  \right) \,. 
 \label{eq:logBE}
\end{equation}
We use a shooting algorithm to solve the equations with an initial condition at high temperatures, and adaptively switch between solving \eqref{eq:BE} and \eqref{eq:logBE} in different regimes, appropriately matching solutions at the boundary. We verify numerically that our solutions do not depend on the initial condition, other than the assumption that $n \ll n_{\rm eq}$ at early times \cite{Berger:2018xyd}.

\begin{figure*}
    \centering
    \includegraphics[width=1\textwidth]{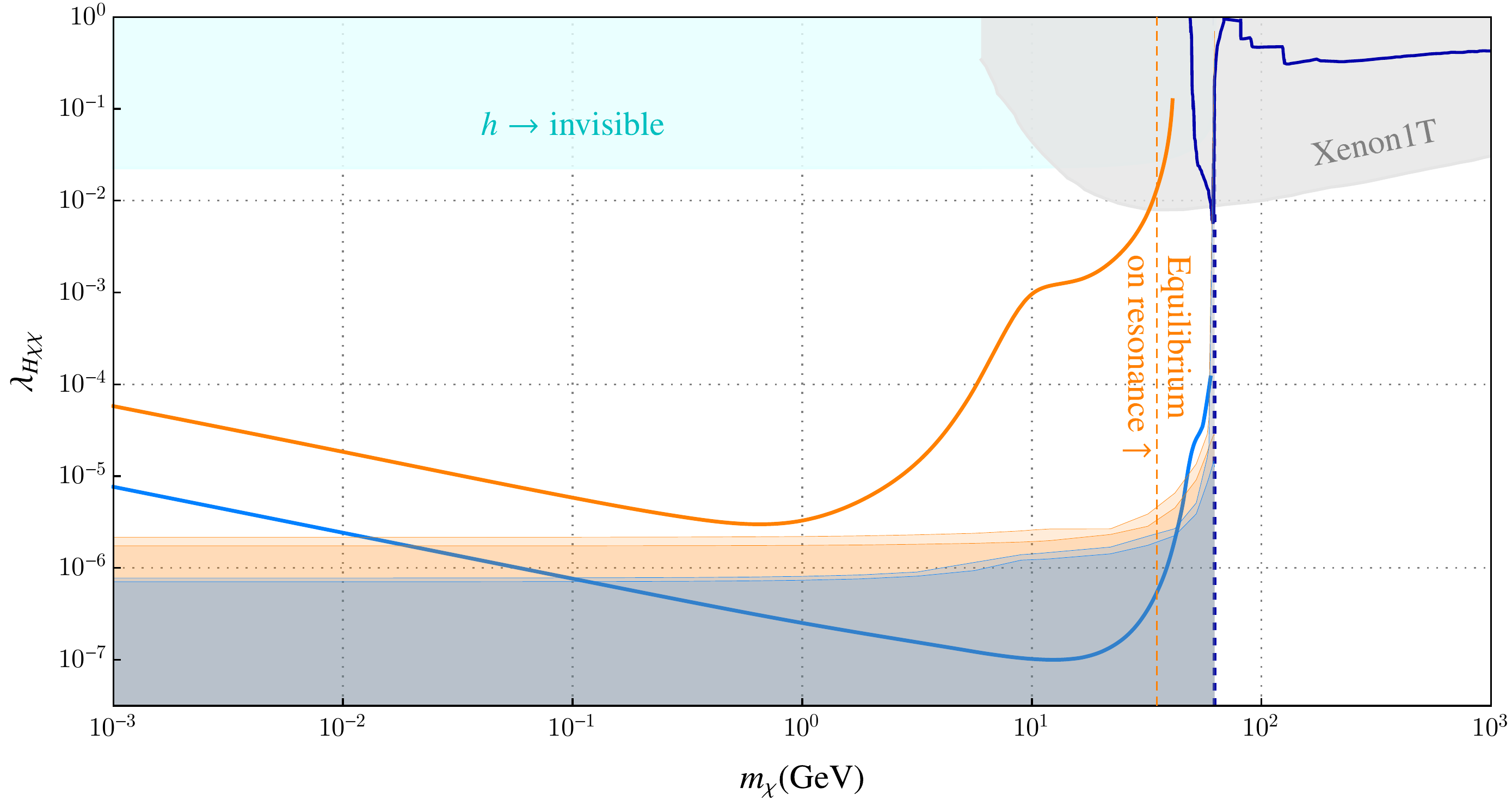}
    \caption{Relic dark matter abundance for a morphon benchmark as in Fig.~\ref{fig:rates} (orange, upper curve) and twice that morphon mass and effective coupling: $m_\phi = 2 \times 10^{-10}$ eV, $\sqrt{\lambda_\phi} = 5 \times 10^{-20} y_{\phi \chi} $  (blue, lower curve), as a function of the zero-temperature mass $m_\chi$ and coupling $\lambda_{H \chi\chi}$. 
    For small dark matter masses, the behavior mirrors that of freeze-in dark matter: the abundance scales with $\lambda_{H\chi\chi}^2$.
    For larger dark matter masses, the resonance temperature shifts downwards, and as a result larger couplings are needed to produce the observed dark matter yield. The first (upper, orange) benchmark model brings the dark matter into equilibrium on the resonance for zero-temperature dark matter masses $\gtrsim 35$ GeV. The curve around $10$~GeV is explained by threshold effects in the Ricci scalar around the resonance temperature, imprinting on $n_{\rm eq}(T)$ and hence affecting the Boltzmann evolution. The second (blue, lower) benchmark does not go in equilibrium in the presented parameter space. 
    The orange and blue shaded regions indicate where conditions \eqref{eq:kineq} and \eqref{eq:masscond} are violated on the resonance (bottom and top boundary for orange and blue respectively) and hence the treatment is not valid. 
    }
    \label{fig:relic}
\end{figure*}

Numerically integrating the Boltzmann equations leads to a result for the dark matter yield $Y\equiv n/s$, where $s$ is the entropy density (with a value today of $s_0 = 2970 \,{\rm cm}^{-3}$). We find solutions where the produced yield matches the measured value $Y=3.55 \times 10^{-10} ({\rm GeV}/m_{\rm DM})$ \cite{Aghanim:2018eyx} at $T_0 = 2.73 $ K. A benchmark solution is shown in Fig.~\ref{fig:yield}, demonstrating the thermal evolution through the resonance at the temperature $T_\text{res}$. At this temperature, a sharp increase in the yield appears, allowing small portal couplings to set the correct relic abundance.

Two benchmark relic abundance curves are given in Fig.~\ref{fig:relic}. 
Unlike freeze-in models, the dominant production channel is not Higgs decay, as the low resonance temperature implies the Higgs number density is very small. The abundance is determined by the annihilation cross section at the resonance temperature. At low  masses, the production therefore scales with $\lambda_{H {\rm dm}}^2$ for a given morphon benchmark. 
At larger  masses, the zero-temperature mass of the dark matter shifts the resonance temperature, and larger couplings are needed to produce the relic abundance. 
For final dark matter masses approaching $m_h/2$, the yield is less affected by the resonance peak. In this region, the dominant effect is the growth in equilibrium number density due to the changing dark matter mass. 
While the effect is less dramatic, the relic abundance can still be produced for smaller couplings, suggesting that this behavior may apply to other models with temperature-varying dark matter mass.

Figure~\ref{fig:relic} also summarizes the dominant constraints. Note that we find the constraints on invisible Higgs decays are stronger than constraints on energy injections during Big Bang nucleosynthesis \cite{Finkbeiner:2011dx} and during or after recombination \cite{Finkbeiner:2011dx,Aghanim:2018eyx} - the usual impediments for light dark matter to be produced while in thermal contact with the SM. Recent work has shown that Lymann-$\alpha$ constraints on the matter power spectrum may be strong in out-of-equilibrium scenarios \cite{Dvorkin:2020xga}. We confirm using the approximate techniques given in ref. \cite{Bae:2017dpt} that for the mass range we consider ($\geq 1$ MeV), dark matter produced from our mechanism is well away from Lyman-$\alpha$ bounds \cite{Garzilli:2019qki,Slatyer:2015jla, Liu:2020wqz}. Direct detection bounds in the light dark matter region complement the Xenon1T constraint, but do not place a strong enough bound on the portal coupling to compete with the bound from invisible Higgs decays and so do not restrict our parameter space~\cite{Barak:2020fql, Bernreuther:2020koj, Aprile:2019xxb, Abdelhameed:2019hmk, Agnese:2018gze, Aprile:2019jmx, Knapen:2017xzo}.

The morphon is light enough ($\sim10^{-10}$ eV) that fifth force constraints should be considered. The morphon, however, only couples directly to $\chi$, which in turn only couples to the SM through the Higgs portal. 
As the morphon is a pseudoscalar, $h-\phi$ mixing is forbidden. The morphon therefore cannot mediate a long-range force between SM particles.
While the pseudoscalar nature of the morphons disallows mixing with the Higgs, production of two morphons in the final state is generally possible. Such production would approximately scale as $\Gamma \sim \lambda_{h \chi}^2 \, y_{\phi \chi}^4 \left(m_f/v_h\right)^2 \left({m_\chi}/{m_h^2}\right)^2 \times n_f$, 
where $f$ is a Standard Model fermion in the plasma. Because of the rapid decrease of $\left(m_{f_i}/v_h\right)^2 n_{f_i}$ with temperature, the perturbative production of morphons quickly becomes unimportant below the resonance. 
Then, the morphon number density produced is bounded by $n_{\phi} \leq n_{\phi,{\rm eq}}(T_{\rm dec})$ with $T_{\rm dec} \gtrsim 1 $~GeV. Since the morphon decouples when relativistic, it redshifts as radiation, and therefore gives a maximal contribution to $N_{\rm eff}$:
    \begin{align}
    \Delta N_{\rm eff} = \frac47 \left( \frac{11}4 \right)^{4/3} \left( \frac{ g_*^s(T_\text{CMB}) }{ g_*^s (T_\text{dec}) } \right)^{4/3} \,,
    \end{align}
where $g_*^s(T)$ is the effective number of entropy degrees of freedom, and $T_\text{dec}$ is the temperature at which the morphon decouples~\cite{Blennow:2012de}. To keep the contribution to $\Delta N_{\rm eff}< 0.3$, we require $g_*^s(T_{\rm dec})\gtrsim 18$, which is safely accommodated by $T_\text{dec} \gtrsim 1$ GeV.

\section{Discussion}
In this work we present a new mechanism for the production of light fermionic dark matter via a Higgs portal. In this mechanism, the mass of the dark matter is set dynamically, through the evolution of a scalar field, and as such there is a temperature $T_{\rm res}$ for which $m_{\rm DM} = m_h/2$. The resulting resonant production allows for an observed relic abundance associated with much smaller portal couplings than in the standard freeze-out scenario. 

The evolution of the dark matter mass requires a careful treatment of thermal averaging. We define three conditions to check the validity of the approach. We note that these conditions do not constitute physical constraints, but rather boundaries outside of which further effects should be taken into account. 

We demonstrate the mechanism in a model with a non-minimally coupled morphon, allowing for a dark matter mass which evolves from heavy to light as the temperature drops.
We note that the mechanism could instead be realized with a rolling morphon field, though the oscillations around its minimum would imply rapid production of dark matter fields below the resonance temperature. 

 \vspace{0.4in}
\section*{Acknowledgments}
The authors thank Tim Cohen, Jeff Dror, Miguel Escudero, Lucien Heurtier, Robert McGehee, David McKeen, Toby Opferkuch, and Katelin Schutz for useful discussions. GE is supported by the U.S. Department of Energy, under grant number DE-SC0011637. RH is supported by the STFC under grant ST/P001246/1. TRIUMF receives federal funding via a contribution agreement with the National Research Council Canada. GE thanks the Berkeley Center for Theoretical Physics and Lawrence Berkeley National Laboratory for their hospitality during the completion of this work. The work of HM was supported by the NSF grant PHY-1915314, by the U.S. DOE Contract DE-AC02-05CH11231, by the JSPS Grant-in-Aid for Scientific Research JP20K03942, MEXT Grant-in-Aid for Transformative Research Areas (A) JP20H05850, JP20A203, and Hamamatsu Photonics, K.K. The works of HM and GW were supported by World Premier International Research Center Initiative (WPI), MEXT, Japan. 

\bibliography{references}

\end{document}